\documentclass[12pt, leqno]{revtex4}
\usepackage{epsfig}
\usepackage{amsmath}
\begin{document}
\title{Localization-Delocalization Transition in a Quantum Dot}
\author{MYUNG-HOON CHUNG
%\footnote{Corresponding author. Tel: +82-041-860-2215; fax: +82-041-863-2648. \\
%~~~~~~~~~~~~$E$-$mail$ $address$: mhchung@wow.hongik.ac.kr (M.H. Chung).},
\footnote{$E$-$mail$ $address$: mhchung@wow.hongik.ac.kr,~
$Permanent$ $Address$: College of Science and Technology, Hong-Ik University,
Chochiwon, Choongnam 339-800, Korea }}

\affiliation{Korea Institute for Advanced Study, Dongdaemun-gu,
Seoul 130-012, Korea}

\maketitle

\section*{Abstract}

A model Hamiltonian is proposed in order to understand the
localization-delocalization transition in a quantum dot, where
there are two gate voltages: top and side. Considering energetically
favorable degrees of freedom only, we achieve a finite dimensional
Hilbert space. As a result, exact diagonalization is
performed to find the ground state energy of the system.
It is the purpose to explain the peculiar pattern of the
electron addition energy measured in the dot of two gate voltages.

$PACS$: 73.23.Hk, 73.20.Jc, 73.20.Dx

%$Keywords$: Quantum Dots; Coulomb blockade; Single electron tunneling; Delocalization

\newpage

%\section{Introduction}

Many-body effects in low dimensional systems have attracted many
interests in both experimental and theoretical point of view.
As one of low dimensional systems, quantum dots are fabricated and
used in order to investigate notorious problems of electron
correlation. Ashoori invented the useful tool called single electron
capacitance spectroscopy\cite{1} to measure electronic properties of quantum dots.
It became possible to measure directly the $N$-electron ground state
energies of quantum levels of a dot as a function of magnetic field\cite{2}.
Further studies on electron addition spectra of quantum dots
showed that there are bunches in electron additions\cite{3}.
This strange electron correlation of bunching was investigated
by intensive theoretical efforts\cite{4,5,6}.

Recently, Ashoori group observed the localization-delocalization
transition\cite{7}, using newly designed dots with two kinds of
gate voltages: top $V_{\mbox{t}}$ and side $V_{\mbox{s}}$. The side
gate voltage plays a crucial role in analyzing the edge state
localization. In this seminal experiment, what they measured is the
dependence of the capacitance peak on $V_{\mbox{t}}$ and
$V_{\mbox{s}}$ in electron additions. They plotted the lines of
$V_{\mbox{t}}$ versus $V_{\mbox{s}}$ for each $N$-electron in the
dot up to about $N=100$. The features of the plot are summarized as
the followings. a)Two kinds of lines are observed: one has a small
slope, and the other is steep. b)For the low densities of
electrons, the spacings between adjacent lines are irregular.
c)However, one observes the general trend of decreasing of the
spacings as more electrons are added. d)The small-slope lines
gradually become more steep as the number of electrons in the dot
is increased. e)The anticrossing takes place when a small-slope
line and a steep line are merging. The interesting observation
connected with the anticrossing is that there are two kinds of
anticrossings: normal and abnormal. The anomalous anticrossing
shows that the chemical potential of $N+1$-electron state is lower
than that of $N$-electron state. In fact, this striking result
shows that the edge localized electrons appear to bind with
electrons in the dot center.

It is the purpose of this Letter to explain the
localization-delocalization transition with a tractable model
Hamiltonian. A reduction of the corresponding Hilbert space is
proposed in order for the system to be calculable. This truncation
is called single level approximation, which is resemblance of the
lowest Landau level approximation in the fractional quantum Hall
effect\cite{8,9}. Up to the single level approximation, we
diagonalize the Hamiltonian exactly, using the Lancz\"{o}s method.
It is found that the spacings between small-slope lines are
attributed by Coulomb blockade. It is understood that the steep
lines are related with localized states. Furthermore, we notice
that the anomalous anticrossings are possible by the quantum
interference. In consequence, all features of the plot described
in the above, except for b), are explained in this Letter.

%\section{Hamiltonian}

In order to study the quantum dot, which is experimentally
investigated in Ref. 7, we consider a model Hamiltonian, which is written
as a function of the side gate voltage $V_{\mbox{s}}$:
\begin{equation}
H(V_{\mbox{s}})=H_{\mbox{ext}}+H_{\mbox{loc}}+H_{\mbox{int}},
\end{equation}
where $H_{\mbox{ext}}$ describes the inside of the dot
relating with extended
states, $H_{\mbox{loc}}$ corresponds to localized states, and $H_{\mbox{int}}$
represents interactions between extended electrons and localized electrons.
The measured top gate voltage $V_{\mbox{t}}$ will be a function of
$V_{\mbox{s}}$ in relation with
the electron addition energies:
$\alpha e \Delta V_{\mbox{t}}(N)=E_{\mbox{g}}(N+1)-2E_{\mbox{g}}(N)+E_{\mbox{g}}(N-1)$,
where $E_{\mbox{g}}(N)$ is the ground state energy of
$H(V_{\mbox{s}})$ for total $N$-electron in the system,
and the parameter $\alpha$ is a geometrical coefficient.

While it is difficult to present $H_{\mbox{loc}}$ in terms of
position and momentum variables, we write the extended state
Hamiltonian $\bar{H}_{\mbox{ext}}$
in the $B$-field as
\begin{equation}
\bar{H}_{\mbox{ext}}=\sum_{i=1}^{N}\{\frac{1}{2m^{\ast}}|\vec{p}_{i}+\frac{e}{c}\vec{A}_{i}|^{2}+
\frac{1}{2}m^{\ast}\omega^{2}|\vec{\rho_{i}}|^{2}-g \frac{e}{m^{\ast}c}\vec{S}_{i}\cdot \vec{B} \}
+\sum_{1 \leq i<j \leq N}\frac{e^{2}}{\kappa |\vec{\rho}_{i}-\vec{\rho}_{j}|},
\end{equation}
where we adopt a two-dimensional pancake-type dot with a confining
potential controlled by $\omega$, and the position of the $i$-th
electron is presented by $\vec{\rho}_{i}$. The $g$-factor is
usually given by a very small value enough to ignore the spin term.
The dielectric constant $\kappa$ is introduced in the Coulomb
interaction. The free part of $\bar{H}_{\mbox{ext}}$ was solved by
Fock\cite{10}. In fact, introducing one-particle creation operators
$c_{nm\sigma}^{\dagger}$, we find the corresponding eigenfunctions:
$\Psi_{nm\sigma}(\rho,\phi)=$
$<\vec{\rho}|c_{nm\sigma}^{\dagger}|0>=$ $\sqrt{2A/\pi
((n-1-|m|)/2)!/((n-1+|m|)/2)!}$ $\exp(-i m
\phi)\exp(-A\rho^{2})(2A\rho^{2})^{|m|/2}$
$L^{|m|}_{(n-1-|m|)/2}(2A\rho^{2})$ $\chi^{\mbox{spin}}_{\sigma}$,
where the principal quantum number $n$ runs as $1$, $2$, $3$,
$\cdots$; the corresponding magnetic quantum number $m$ is given by
$-(n-1),~-(n-3),~\cdots,~(n-3),~(n-1)$; the spin index
$\sigma=\uparrow,\downarrow$; and the short hand notation
$A=m^{\ast}\bar{\omega}/2\hbar$,
$\bar{\omega}=\sqrt{\omega^{2}+\frac{1}{4}\omega_{\mbox{c}}^{2}}$,
$\omega_{\mbox{c}}=eB/m^{\ast}c$. Here, following the normalization
of Arfken\cite{11}, we note the associated Laguerre polynomials:
$L_{n}^{k}(x)=$$\sum_{m=0}^{n}(-1)^{m}$
$(n+k)!/\{(n-m)!(k+m)!m!\}x^{m}\equiv$
$\sum_{m=0}^{n}A_{n}^{k}(m)x^{m}$,
where the fractional numbers of $A_{n}^{k}(m)$ will be used to present the
coefficients for the Coulomb interaction in the formalism of
second quantization.

It is a straightforward process to obtain the second quantized
Hamiltonian. With complicated coefficients for the Coulomb
interaction, the second quantized Hamiltonian will be written as
the usual form in terms of $c_{nm\sigma}^{\dagger}$ and
$c_{nm\sigma}$. For this full Hamiltonian, the dimension of the
corresponding Hilbert space is infinite. It is impossible to
calculate an exact ground state energy. Thus, we need truncation.
In the case of a small $B$-field, the principal quantum number $n$
plays the role of distinguishing shells. When we consider $N$
electrons in the dot, we note that the electrons occupy from lower
energy states. There will be the biggest value of $n$ in this
situation. We can divide the number of electrons as
$N=N_{\mbox{core}}+N_{\mbox{shell}}$, where
$N_{\mbox{core}}=(n-1)n$, and $N_{\mbox{shell}}=0,~1,\cdots,~2n$.
In this Letter, reducing the degrees of freedom, we ignore detailed
interactions between principal quantum levels, and also neglect
higher energy states. This is called single level approximation.
Like the case of Ref. 7, we simply let the $B$-field zero from now
on, hence $\omega_{\mbox{c}}=0$ and $\bar{\omega}=\omega$.
Roughly taking care of the interaction between the core and the shell
electrons, we introduce parameters $C(N_{\mbox{core}})$. In
consequence, the truncated extended state Hamiltonian is written as
\begin{eqnarray}
H_{\mbox{ext}}&=&E(N_{\mbox{core}})+C(N_{\mbox{core}})N_{\mbox{op}}+\hbar \omega n N_{\mbox{op}} \\
&+&\sum_{-(n-1)\le k,l+m,l,k+m \le (n-1)} V_{n}(k,l,m)\sum_{\sigma,\sigma^{\prime}}
c_{nk\sigma}^{\dagger}c_{nl+m\sigma^{\prime}}^{\dagger}c_{nl\sigma^{\prime}}c_{nk+m\sigma},
\nonumber
\end{eqnarray}
where
$N_{\mbox{op}}=\sum_{m\sigma}c_{nm\sigma}^{\dagger}c_{nm\sigma}$,
and the zero-point energy $E(N_{\mbox{core}})$ effectively
describes the interaction among the core electrons. Now, the
dimension of the corresponding Hilbert space is finite, in fact,
the number of different ways in taking $N_{\mbox{shell}}$ out of
$2n$. Thus, it is calculable. The values of $E(N_{\mbox{core}})$
and $C(N_{\mbox{core}})$ will be determined later. Using the simple
formula:
%\begin{equation}
$1/|\vec{\rho}_{2}-\vec{\rho}_{1}|=
\sum_{m\in
\mbox{Z}}$
$\exp(im(\phi_{1}-\phi_{2}))\int_{0}^{\infty}dkJ_{m}(k\rho_{1})J_{m}(k\rho_{2})$,
%\end{equation}
where $J_{m}(x)$ is the Bessel function, we calculate the
coefficients $V_{n}(k,l,m)$ of the Coulomb interaction in the level
of $n$:
\begin{eqnarray}
V_{n}(k,l,m)=\frac{e^{2}}{2\kappa}\sqrt{A}
\sqrt{\frac{\frac{n-1-|l+m|}{2}!\frac{n-1-|l|}{2}!\frac{n-1-|k+m|}{2}!\frac{n-1-|k|}{2}!}
{\frac{n-1+|l+m|}{2}!\frac{n-1+|l|}{2}!\frac{n-1+|k+m|}{2}!\frac{n-1+|k|}{2}!}} \\ \nonumber
\times
\sum_{k_{1}=0}^{\frac{n-1-|l+m|}{2}}
\sum_{k_{2}=0}^{\frac{n-1-|l|}{2}}
\sum_{k_{3}=0}^{\frac{n-1-|k+m|}{2}}
\sum_{k_{4}=0}^{\frac{n-1-|k|}{2}}
A^{|l+m|}_{\frac{n-1-|l+m|}{2}}(k_{1})
A^{|l|}_{\frac{n-1-|l|}{2}}(k_{2})
A^{|k+m|}_{\frac{n-1-|k+m|}{2}}(k_{3})
A^{|k|}_{\frac{n-1-|k|}{2}}(k_{4}) \\ \nonumber
\times
(k_{1}+k_{2}+\frac{|l+m|+|l|-|m|}{2})!
(k_{3}+k_{4}+\frac{|k+m|+|k|-|m|}{2})!  \\ \nonumber
\times
\sum_{l_{1}=0}^{k_{1}+k_{2}+\frac{|l+m|+|l|-|m|}{2}}
\sum_{l_{2}=0}^{k_{3}+k_{4}+\frac{|k+m|+|k|-|m|}{2}}
A^{|m|}_{k_{1}+k_{2}+\frac{|l+m|+|l|-|m|}{2}}(l_{1})
A^{|m|}_{k_{3}+k_{4}+\frac{|k+m|+|k|-|m|}{2}}(l_{2})\\ \nonumber
\times
\frac{\Gamma(|m|+l_{1}+l_{2}+\frac{1}{2})}{2^{|m|+l_{1}+l_{2}}}.
\end{eqnarray}
Note that the overall factor is proportional to $\sqrt{\omega}$ as
$e^{2}\sqrt{A}/2\kappa \equiv V=
V_{0}\sqrt{\hbar \omega / 1 \mbox{meV}}$.
Comparing $\sqrt{\omega}$ with $\hbar \omega$, which is the energy difference
between principal quantum levels, we find
that our single level approximation is the more valid for the larger
value of $\omega$, that is, the stronger confinement.

Turning our attention to localized electrons, we consider only the
case of a single localized state without loss of generality.
We guess the Hamiltonian $H_{\mbox{loc}}$ with the creation operator
$d^{\dagger}_{\sigma}$ for the localized electron as
\begin{equation}
H_{\mbox{loc}}=\epsilon (n_{\uparrow}+n_{\downarrow})+Un_{\uparrow}n_{\downarrow},
\end{equation}
where $n_{\sigma}=d_{\sigma}^{\dagger}d_{\sigma}$.
It seems that two localized electrons at the same site are most likely
feel a large effect of repulsion. Thus, the value of $U$ should be
large. The energy value of $2\epsilon + U$ is
energetically unfavorable. Thus,
it is enough to consider only two cases, 0 or $\epsilon$, for the
energy of $H_{\mbox{loc}}$.

The interaction Hamiltonian between extended and localized
electrons is written as
\begin{equation}
H_{\mbox{int}}=\sum_{n,m,\sigma}\{\lambda_{nm}c_{nm\sigma}^{\dagger}c_{nm\sigma}
(n_{\uparrow}+n_{\downarrow})+
t_{nm}c_{nm\sigma}^{\dagger}d_{\sigma}
+t_{nm}^{\ast}d_{\sigma}^{\dagger}c_{nm\sigma}\}.
\end{equation}
Here we consider the direct Coulomb interaction and the tunneling effect.
Since the localized electron wave function
$<\vec{\rho}|d_{\sigma}^{\dagger}|0>$ is not known, we can not
calculate $\lambda_{nm}$, nor $t_{nm}$.

We have introduced the Hamiltonian of the system, $H(V_{\mbox{s}})$.
Our task is now to find the ground state energy
of $H(V_{\mbox{s}})$. The
strategy for this is to calculate the ground state energy of $H_{\mbox{ext}}+H_{\mbox{loc}}$
first, and to use the Rayleigh-Schr\"{o}dinger perturbation theory with respect to
$H_{\mbox{int}}$. And then, we make connection between $V_{\mbox{t}}$
and $V_{\mbox{s}}$, using the electron addition energy.

We have already determined the energy of $H_{\mbox{loc}}$
trivially. We consider $H_{\mbox{ext}}$ for a corresponding
principal quantum number $n$ with the single level approximation.
We write the ground state energy of $H_{\mbox{ext}}$ in Eq. (3) as
\begin{equation}
E(N)=E(N_{\mbox{core}})+C(N_{\mbox{core}})N_{\mbox{shell}}+\hbar \omega n N_{\mbox{shell}}
+E_{\mbox{c}}(N),
\end{equation}
where $E_{\mbox{c}}(N)$ is the Coulomb correlation energy, which
plays an essential role in the electron addition energy. In the
single level approximation, it is obvious to notice
$E_{\mbox{c}}(N_{\mbox{core}}+1)=0$. We calculate the correlation
energy $E_{\mbox{c}}(N)$ up to $N=170$ by using the Lancz\"{o}s
method, for instance, $E_{\mbox{c}}(1)=0.0$, $E_{\mbox{c}}(2)=1.77245V$,
$E_{\mbox{c}}(3)=0.0$, $E_{\mbox{c}}(4)=0.88622V$, $\cdots$,
$E_{\mbox{c}}(156)=129.724V$, $\cdots$, and $E_{\mbox{c}}(170)=39.8113V$.

Using the ground state energy of $H_{\mbox{ext}}$ in Eq. (7), we calculate
the electron addition energy $E(N+1)-2E(N)+E(N-1)$ as
\begin{equation}
\begin{cases}
\hbar \omega + \Delta(n) & \text{for } N=n(n+1), \\
E_{\mbox{c}}(n(n+1)+2) & \text{for } N=n(n+1)+1, \\
E_{\mbox{c}}(N+1)-2E_{\mbox{c}}(N)+E_{\mbox{c}}(N-1) & \text{otherwise,}
\end{cases}
\end{equation}
where $\Delta(n)=C(n(n+1))-C((n-1)n) -
E_{\mbox{c}}(n(n+1))+E_{\mbox{c}}(n(n+1)-1)$. Since it is
numerically shown that the value of
$E_{\mbox{c}}(n(n+1)-1)-E_{\mbox{c}}(n(n+1))$ contains a negative
factor proportional to $n$, the parameter $C(n(n+1))$ must cancel
the factor by subtraction so that $\hbar \omega + \Delta(n)$ is
positive for all $n$ to follow the concept of Coulomb blockade. The
parameters $C(n(n+1))$ are determined by $\Delta(n)$, which should
be properly chosen to satisfy experimental data. All values of
$E(N_{\mbox{core}})$ and $C(N_{\mbox{core}})$ are determined
recursively from $E(0)=C(0)=0$.

We write two candidates $|\Psi_{N+1}>$ and $d^{\dagger}_{\sigma}|\Psi_{N}>$
for the ground state of the Hamiltonian $H_{\mbox{ext}}+H_{\mbox{loc}}$
with $N+1$ electrons as
$(H_{\mbox{ext}}+H_{\mbox{loc}})|\Psi_{N+1}>=$
$E(N+1)|\Psi_{N+1}>$, and
$(H_{\mbox{ext}}+H_{\mbox{loc}})d^{\dagger}_{\sigma}|\Psi_{N}>=$
$(E(N)+\epsilon)d^{\dagger}_{\sigma}|\Psi_{N}>$.
Considering the total Hamiltonian of Eq. (1) now, we calculate the
correction energy of the first order perturbation as
$<\Psi_{N+1}|H_{\mbox{int}}|\Psi_{N+1}>=0$
and
$<\Psi_{N}|d_{\sigma}H_{\mbox{int}}d^{\dagger}_{\sigma}|\Psi_{N}>=$
$<\Psi_{N}|\sum_{n,m,\sigma}\lambda_{nm}c^{\dagger}_{nm\sigma}c_{nm\sigma}|\Psi_{N}>$
$\equiv \lambda(N)$.
Note that the ground state energy of the system is $E(N+1)$ or
$E(N)+\epsilon+\lambda(N)$. If these two values
are almost same, then this is the degenerate case and
we should diagonalize a matrix in perturbation.
With the two degenerate states $d^{\dagger}_{\sigma}|\Psi_{N}>$ and
$|\Psi_{N+1}>$, we find the relevant Hamiltonian
$H_{\mbox{cross}}$:
\begin{equation}
H_{\mbox{cross}}= \left( \begin{array}{cc}
         E(N)+\epsilon+\lambda (N) & t^{\ast}(N) \\
         t(N) & E(N+1)
        \end{array}   \right),
\end{equation}
where $<\Psi_{N+1}|\sum_{n,m,\sigma}t_{nm}c^{\dagger}_{nm\sigma}|\Psi_{N}>
\equiv t(N)$. We find that the degeneracy is removed by $t(N)$, and the ground
state energy is given by
$\frac{1}{2}\{E(N+1)+E(N)+\epsilon+\lambda (N)$
$-\sqrt{(E(N+1)-E(N)-\epsilon-\lambda (N))
^{2}+4|t(N)|^{2}}\}$
$\equiv$
$E_{\mbox{cross}}(N+1)$.
In consequence, the ground state energy of the system $E_{\mbox{g}}(N+1)$
up to the first order perturbation with respect to $H_{\mbox{int}}$
is given by $\mbox{min}\{E(N)+\epsilon+\lambda(N),E(N+1)\}$ or $E_{\mbox{cross}}(N+1)$
if $E(N)+\epsilon+\lambda(N) \approx E(N+1)$.
This result of $E_{\mbox{g}}(N+1)$ and also $E_{\mbox{g}}(N)$
will be used in the below.

Since a line measured in the experiment presents the event of
single electron oscillation between the dot and the contact, it is
appropriate to use the notation of $\frac{1}{2}$ in
$eV_{\mbox{t}}(N+\frac{1}{2})$. Introducing the offset value
$eV_{\mbox{t}}(\frac{1}{2})$, which is the first line in the plot
of $V_{\mbox{t}}$ versus $V_{\mbox{s}}$, we find
$eV_{\mbox{t}}(N+\frac{1}{2})=$ $eV_{\mbox{t}}(\frac{1}{2})+$
$\sum_{i=1}^{N}e\Delta V_{\mbox{t}}(i)=$
$eV_{\mbox{t}}(\frac{1}{2})+$
$\frac{1}{\alpha}\{E_{\mbox{g}}(N+1)-E_{\mbox{g}}(N)-$
$E_{\mbox{g}}(1)+E_{\mbox{g}}(0)\}$. Note that the energy
differences $D(N+1)\equiv E(N+1)-E(N)-\lambda(N)$ satisfy the
inequality of $D(N) < D(N+1)$ for all $N$ with small $\lambda(N)$.
We find that $E_{\mbox{g}}(N+1)-E_{\mbox{g}}(N)$ in
$eV_{\mbox{t}}(N+\frac{1}{2})$ is given by one of the five expressions
according to $\epsilon$:
\begin{equation}
\begin{cases}
D(N)+\lambda(N) & \text{for } \epsilon < D(N), \\
E(N)+\epsilon+\lambda(N)-E_{\mbox{cross}}(N)& \text{for } \epsilon \approx D(N), \\
\epsilon+\lambda(N) & \text{for } D(N)<\epsilon<D(N+1), \\
E_{\mbox{cross}}(N+1)-E(N) & \text{for } \epsilon \approx D(N+1), \\
D(N+1)+\lambda(N) & \text{for } \epsilon > D(N+1). \\
\end{cases}
\end{equation}
As far as parameters in the Hamiltonian are functions of
$V_{\mbox{s}}$, $E_{\mbox{g}}(N)$ is also a function of
$V_{\mbox{s}}$. Hence, we have connected $V_{\mbox{s}}$ to
$V_{\mbox{t}}(N+\frac{1}{2})$. In Ref. 7, observing the large
capacitance shows that the localized states exist at the periphery
of the dot. The energy value of the localized state $\epsilon$ is
much affected by $V_{\mbox{s}}$. Furthermore, the side gate voltage
$V_{\mbox{s}}$ effectively changes the dot confining potential.
Thus, we assume the dependence of the side gate voltage on
the parameters as
\begin{equation}
\epsilon=\epsilon_{0}+\beta e |V_{\mbox{s}}|,~~
\hbar\omega=\hbar\omega_{0}+\gamma e |V_{\mbox{s}}|,~~
eV_{\mbox{t}}(\frac{1}{2})=eV_{\mbox{t}0}+\delta e |V_{\mbox{s}}|.
\end{equation}
Summing up, we
plot $V_{\mbox{t}}(N+\frac{1}{2})$ versus $V_{\mbox{s}}$ in
Fig. 1, using Eqs. (7-11). The plot is the main result of this work.
We can clearly see the single line of localization-delocalization
transition, which has a relatively steep slope.
The spacings between the adjacent lines are gradually decreasing.
The slopes of lines are gradually increasing. The unexpected
relatively large spacings appearing periodically is attributed by
our limitation of the single level approximation.
The spacings between adjacent lines, the slope of the special
single steep line, the change of the slopes of small-slope lines, and the
slope of the first line are controlled by the values of
$\alpha,~\beta,~\gamma,~\delta$, respectively.
The starting points of the steep line and the first line are
determined by $\epsilon_{0}$ and $eV_{\mbox{t}0}$, respectively.
Our plot of Fig. 1 shows the regular spacings between small-slope lines even
for low electron densities. This is not in agreement with experimental results.
It seems that the extended state Hamiltonian $H_{\mbox{ext}}$ is only valid for
high electron densities.

We can notice the anticrossings between the steep line and
the small-slope lines. From the plot and Eq. (10), in order to
study the anomaly of anticrossing, we should compare
$V_{\mbox{t}}(N+1+\frac{1}{2})$ in the region of $\epsilon <D(N+1)$
with $V_{\mbox{t}}(N+\frac{1}{2})$ in the region of $\epsilon >
D(N+1)$. We find that, if $E(N+1)+\lambda(N+1)-E(N)-\lambda(N)$ is
equal to or greater than $E(N+1)-E(N)$, it is a normal
anticrossing, and if otherwise, it is abnormal. In fact, we note
$\lambda(N+1)<\lambda(N)$ for the cases of anomalous anticrossings.
In the classical point of view, it is expected that the value of
$\lambda(N)$ is always increasing as $N$ becomes bigger. However,
because of the quantum interference in the ground state of the
system, it seems that the expectation value of $\lambda(N+1)$ can
be smaller than $\lambda(N)$.

In conclusion, we have considered a model Hamiltonian containing
interactions between extended electrons and localized
electrons. Coulomb blockade is found in energy calculation with
only the extended state Hamiltonian. The general trends of the plot
$V_{\mbox{t}}$ versus $V_{\mbox{s}}$ obtained by our theoretical study
coincide with the experimental data. Including the localized state
Hamiltonian, we explain the steep line observed in experiment. The
possibility of the anomalous anticrossing is due to the interaction
between extended and localized electrons. Our single level
approximation introduced in this Letter is not applicable to the
case of low electron densities. Perhaps, full calculation with more
detailed Hamiltonian would be useful to explain the irregular
spacings of the addition energies for low electron densities.

%\section*{Acknowledgment}

The author is grateful to Professor C. K. Kim for introduction to
quantum dots.

\begin{figure}
\begin{center}
%\epsfbox{figure1.eps}
%\epsfig{file=figure1.eps,%
\epsfig{file=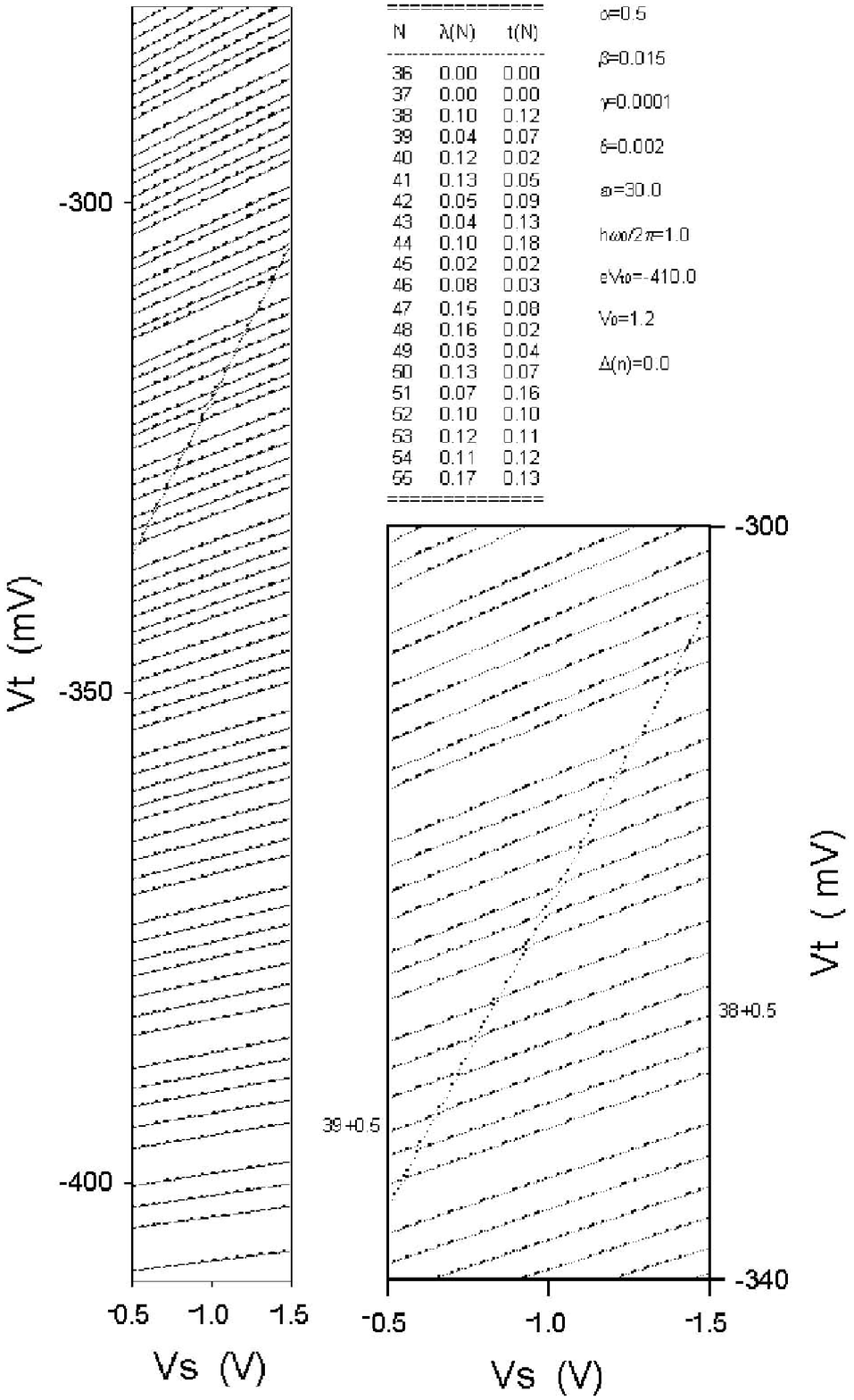,%
        height=18cm}
\end{center}
\caption {The plot of the electron addition energy versus the side gate
voltage is shown. Zoom-in to the part near
$V_{\mbox{t}}=-320\mbox{meV}$ is shown in the right. All quantities
of energy dimension written in the right-top have the unit of
$\mbox{meV}$. The dimensionless parameters
$\alpha,~\beta,~\gamma,~\delta$ are chosen in order to obtain the
similar feature of figure 1 (B) presented in Ref. 7. The values of
$\Delta(n)$ related with $C(N_{\mbox{core}})$ is chosen simply as
zero for all. The values of $\lambda(N)$ and $t(N)$ from $N=36$ to
$55$ are arbitrarily given. A typical anomalous anticrossing takes
place with $\lambda(39)<\lambda(38)$.}
\end{figure}

\end{document}